# *The Problem of Big Bang Matter vs. AntiMatter Symmetry*

*Roger Ellman*

Abstract

The result of the spherically symmetrical Big Bang had to be equal amounts of matter and antimatter with the expectation of their mutual annihilation. The favored explanation for that not happening is that the original symmetry was skewed in favor of matter and the universe is now all matter, the antimatter having annihilated with an equal amount of matter.

That skewed balance conflicts with a purely symmetrical Big Bang and is difficult to justify. Current investigations seek an innate violation of matter / antimatter symmetry sufficient to do so.

An alternative maintaining the original symmetry is presented. The logic and mechanism of mutual annihilation is analyzed and shows that a total mutual annihilation of original matter and antimatter could not have occurred. Our present universe must contain equal amounts of both forms of matter between some particles of which mutual annihilations can occur at a modest rate.

Current indication of detection of cosmic matter / antimatter mutual annihilations is Gamma Ray Bursts [GRB's]. However, the conviction that the universe is now all matter with no antimatter has left that possibility rejected and uninvestigated and left standing the massive supernovae core collapse hypothesis for GRB's.

It has recently been reported [4] that the rate of GRB's increases with red shift $z$ over the range $z = 0 - 4$ as about $(1 + z)^{1.5}$. That is, the indication is that the rate increases significantly with time into the past at least back to the time corresponding to $z = 4$ [and probably back to the Big Bang].

That finding is inconsistent with the massive supernovae core collapse hypothesis for GRB's and supports GRB's being cosmic matter / antimatter mutual annihilations.

Roger Ellman, The-Origin Foundation, Inc.
  320 Gemma Circle, Santa Rosa, CA 95404, USA
  RogerEllman@The-Origin.org
  http://www.The-Origin.org

# *The Problem of Big Bang Matter vs. AntiMatter Symmetry*

*Roger Ellman*

## *1 - Background of the Problem*

It is generally deemed, and reasonably so, that the Big Bang had to be largely symmetrical and exhibit a smooth spherical uniformity in the pattern of particles, energy, and radiation emitted outward in all directions from the singularity source. That would also apply to the emitted particles versus their antiparticles and would imply that the Big Bang should have resulted in equal amounts of matter and antimatter, for which the expectation would be their complete and almost instantaneous mutual annihilation.

On the other hand, because a total mutual annihilation did not take place, as evidenced by our and our universe's existence, the general cosmological position currently favored holds that the original symmetry was slightly skewed in favor of matter, that the universe is now all matter, all original antimatter having been annihilated with an equal amount of original matter.

That skewed balance conflicts with a purely symmetrical Big Bang. Current experiments process seek to detect an innate violation of matter / antimatter symmetry sufficient to justify the skew.

The following development presents an alternative showing that a total mutual annihilation of equal amounts of original matter and antimatter could not have occurred; that, rather, while a moderate amount of initial matter / antimatter mutual annihilations may have taken place our present universe contains the remaining matter and antimatter in equal amounts, between some particles of which further mutual annihilations still occur at a modest rate.

## *2 - Conditions Affecting Matter / AntiMatter Mutual Annihilation*

The first issue to be investigated is the necessary conditions for a matter / antimatter annihilation to take place: how close must the particle and its antiparticle be and for how long must they remain in such sufficiently intimate contact ?

In addition to those two factors there is the more obvious requirement that the two particles involved be true antiparticles of each other [for example, a proton and an antiproton or an electron and a positron, but not a proton and a positron nor a proton and an electron]. Furthermore in general, particle / antiparticle annihilations are relatively unlikely between electrically neutral particles [for example, a neutron and an antineutron] because the only effects tending to bring the two together are their very weak gravitational attraction or chance encounter.

### *The Closeness Criterion*

Indication of how close the two participating particles must be for their annihilation to take place can be found from the decay of a free neutron [not one that is part of an atomic nucleus] into a proton and an electron, a natural process with a mean lifetime before decay of about 881.5 seconds. For the neutron decay to be successful the proton and electron product particles must derive from the parent neutron not only their rest masses but also sufficient kinetic energy so that they are at escape velocity relative to each other, else they would be attracted back together and recombine. [One can neglect the also emitted electron anti-neutrino which is of zero or negligible mass.]

The escape velocity of the two particles is, at first consideration, an awkward problem because the separation distance of the two particles, which appears in the denominator of the expression for their Coulomb attraction, would seem to be able to be as small as zero. That is, at first consideration the escape velocity required is infinite. But, since infinite escape velocity is impossible yet the escape occurs, then the starting point, the minimum separation distance that can occur must be greater than zero. In other words, the neutron decay products, a proton and an electron, exist as such only when separated by some minimum separation distance, $s$, and their state at lesser separation distances appears as their parent neutron.

Therefore, since if the proton and the electron are separated by less than that minimum distance they do not exist as proton and electron but rather as the neutron, and at separation distances greater than that minimum they are the pair of separate particles, then that separation distance is a measure of how close a proton and an electron must be to unite into a neutron and is indicative of the spacing at which a particle and its antiparticle mutually annihilate.

The point is that the excess of the mass of the neutron over that of a proton plus that of an electron must supply the proton and electron relativistic kinetic masses needed to escape the decaying neutron or, alternatively, must derive from the kinetic mass brought to the neutron in the proton and electron being accelerated from infinitely apart to joining in a neutron. The detailed analysis and relativistic calculations can be found in *A New Look at the Neutron and the Lamb Shift*[1] The results are as follows.

*(1)*   – The escape velocities:

$$v_e = 275,370,263. \text{ meters per second}$$
$$= 0.918,536,33 \cdot c$$
$$v_p = 379,350.6975 \text{ meters per second}$$
$$= 0.001,265,378 \cdot c$$

– The minimum separation distance:

$$S = 1.3 \cdot 10^{-15} \text{ meters}$$

where the precision for this separation distance is limited by the precision of Lamb Shift data as discussed in the above referenced paper.

Some years ago experiments involving measurement of the scattering of charged particles by atomic nuclei, yielded an empirical formula for the approximate value of the radius of an atomic nucleus to be

*(2)*   Radius = $[1.2 \cdot 10^{-15}] \cdot$ [Atomic Mass Number] meters

which formula would indicate that the radius of the proton as a Hydrogen nucleus (atomic mass number $A = 1$) is about $1.2 \cdot 10^{-15}$ meters.

The mass of the proton can be expressed as an equivalent energy, $W_p = m_p \cdot c^2$, and that as an equivalent frequency, $f_p = m_p \cdot c^2 / h$, or as an equivalent wavelength, $\lambda_p = c / f = h / m_p \cdot c$. That wavelength (not a "matter wavelength") for the proton is

*(3)*   $\lambda_p = 1.321,410,0 \cdot 10^{-15}$ meters

quite near to the empirical value for the proton radius from equation *(2)* and the separation distance, $s$, of equation *(1)*. Thus the separation distance boundary between a proton and an electron as separate particles versus combined into a neutron is about *1* proton radius, the equivalent wavelength for the proton mass per equation *(3)*.

Then for a proton and an antiproton the boundary between their being the two separate particles and their mutually annihilating is a proton radius, a separation distance of $S_p = \lambda_p = 1.321,410,0 \cdot 10^{-15}$ *meters*. At that boundary if their velocities have a sufficient net component directly toward each other [per the time criterion, below] they would seem to be able, and likely, to mutually annihilate, and otherwise the annihilation would seem not possible.

Similarly, the mass of the electron or the positron can be expressed as the equivalent energy, $W_e = m_e \cdot c^2$, and that as its equivalent frequency, $f_e = m_e \cdot c^2 / h$, or equivalent wavelength, $\lambda_e = c/f = h/m_e \cdot c$. That wavelength (not a "matter wavelength") for the electron / positron is

(4)     $\lambda_e = 2.426,310,6 \cdot 10^{-12}$ meters.

Then for an electron and a positron the boundary between their being the two separate particles and their mutually annihilating is a separation distance of $S_e = \lambda_e = 2.426,310,6 \cdot 10^{-12}$ meters. At that boundary if their velocities have a sufficient net component directly toward each other [per the time criterion, below] they would seem to be able, and likely, to mutually annihilate, and otherwise the annihilation would seem not possible.

Then, what is that sufficient net velocity ?

### *The Time Criterion*

The mutual annihilation of a particle and its antiparticle is symbolized as in the following example for a proton and an antiproton.

(5)     $_1p^1 + {}_{-1}p^1 \Rightarrow \gamma + \gamma$     where γ is a gamma photon

In the present case of a proton and an antiproton the mass of each of the protons is converted into the energy of the related γ photon. The frequency and period of each of those two photons is as follows.

(6)     $f_{\gamma p} = m_p \cdot c^2 / h$

$T_{\gamma p} = 1/f_{\gamma p} = h/[m_p \cdot c^2] = 4.407,749,3 \cdot 10^{-24}$ seconds

In communications theory it is shown that a sinusoidal oscillatory signal must be sampled at least twice per cycle for the signal to be correctly represented. That is, two independent datums are required so as to determine the value of the oscillation's two absolute parameters, its amplitude and its frequency. [It's phase is relative, not absolute.] That implies that the time duration of a proton / antiproton mutual annihilation must be the period of each of the resulting photons.

(7)     $\Delta t_{proton \ / \ antiproton} = T_{\gamma p} = 4.407,749,3 \cdot 10^{-24}$ seconds

Similarly for an electron / positron mutual annihilation, the time duration would be

(8)     $\Delta t_{electron \ / \ positron} = T_{\gamma e} = 8.093,301,0 \cdot 10^{-21}$ seconds.

While those are very brief times they are not instantaneous.

In the case of a particle and its antiparticle coming together from significantly far apart, the particles will have accumulated significant velocity toward each other by the time they arrive at separation distance $S$ because of having been accelerated by their mutual Coulomb attraction. However, the situation was different for the Big Bang.

The number of particles resulting from the original Big Bang is estimated to have been about $10^{85}$ **2**, and those particles emerged on paths that were initially radially outward. The event was overall spherically symmetrical on the large scale, but at the local particle level perfect symmetry was impossible because of the nature of finite particles versus a smooth non-particulate substance. Initially all of the particles were on divergent paths although for two adjacent particles the amount of the divergence was minute.

For a proton and an adjacent antiproton in the Big Bang to be separate [not annihilated] at the instant of being projected outward in the Big Bang, they had to be separated by at least the above-developed $S_p = 1.321,410,0 \cdot 10^{-15}$ meters. For them to annihilate their Coulomb attraction must accelerate them into co-locating in the required time frame starting from their initially having zero velocity toward each other. [Actually they would have had non-zero but minute velocities away from each other because each follows its own outward radial path.] The issue is whether the Coulomb attraction can accelerate the two particles to the point of co-locating within the time frame of equation *(7)* [or equation *(8)* for an electron / positron case].

If, for example, for their mutual annihilation, the proton or the antiproton is to travel <u>at constant velocity</u> its half of the separation distance, $\tfrac{1}{2} \cdot S_p$, in time $T_{\gamma p}$, so as to be co-located with its antiparticle at the end of that time, it would require a speed of

$$(9) \quad v_p = \frac{\tfrac{1}{2} \cdot S_p}{T_{\gamma p}} = 0.5 \cdot c \quad \text{[half light speed]}$$

and if the electron or the positron, for their mutual annihilation, is to travel its half of the separation distance, $S_e$, in time $\tfrac{1}{2} \cdot T_{\gamma e}$ <u>at constant velocity</u> it would require a speed of

$$(10) \quad v_e = \frac{\tfrac{1}{2} \cdot S_e}{T_{\gamma e}} = 0.5 \cdot c \quad \text{[half light speed]}.$$

The achieving of that speed, if even only by the very end of the extremely short time period of the acceleration and travel, $10^{-21}$ seconds or less, would be difficult. The particles moving continuously at that <u>constant velocity</u> throughout their travel from separated to co-located is impossible in that they commence their travel of distance $S$ from essentially zero velocity toward each other.

Furthermore, the analysis of the Coulomb interaction at close separation distances presented in *A New Look at the Neutron and the Lamb Shift*[1] shows that the attraction weakens drastically at close quarters per Figure 1, below, reproduced from that paper. [The figure shows the form of the reduction in the Coulomb attraction as a function of the charge separation radial distance relative to a proton mass equivalent wavelength, $\lambda_p$.]

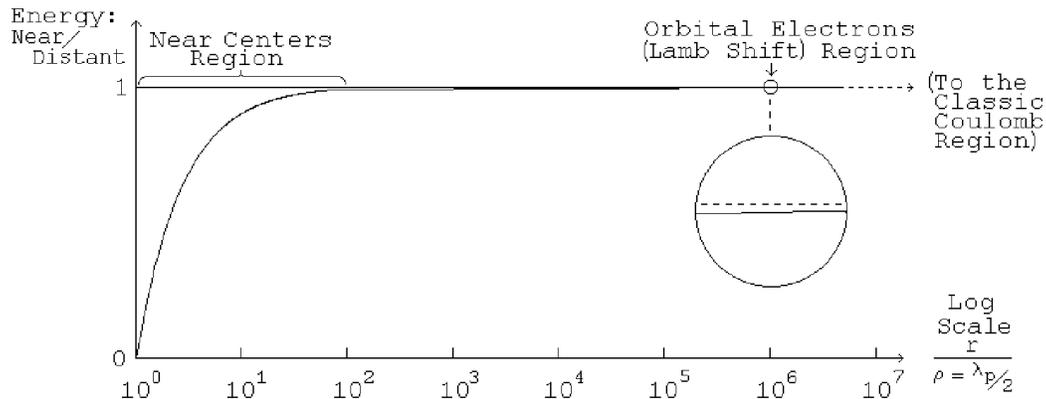

*Figure 1*
*Coulomb Effect <u>Reduction Factor</u> When Charges Are Near to Each Other*

The boundary-like significance of the Separation Distance, that when the two particles are separated by that distance or more they appear as the separate independent particles but when closer than that distance they appear as annihilated, must be taken into account. That would tend to indicate that the portion of the above figure for the region where $r < \lambda_p$ [or $r < \lambda_e$ for the electron / positron case] is probably inapplicable. What happens in that region would appear to be somewhat indeterminate so far as our knowledge can go, but is certainly not the usual Coulomb effect.

Finally, <u>decisively</u>, the posited particle and its antiparticle, emerging from the Big Bang, with spacing adjacent to each other as closely as possible, and on radially outward paths, were not alone. They were surrounded by a more or less uniform, symmetrical, large group of like particles and antiparticles. Any Coulomb tendency to unite the posited particle pair was largely offset by the similar tendency of each member to unite with the adjacent particle on its other side. The net Coulomb action on a specific particle or antiparticle was certainly insufficient to produce enough acceleration to enable the particle to transit its half of the Separation Distance in the required gamma photon period.

In summary:

- adjacent Big Bang product particles and their antiparticles,

- initially spaced optimally [as closely as possible yet independently separate]

- traveling outward at near light speed on essentially parallel paths [actually minutely diverging paths],

- are unable to accelerate toward each other, from essentially zero initial such velocity, quickly enough for their annihilation to produce the known actual gamma photons that would have to result from their mutual annihilation.

- that is, they cannot travel to the point of annihilation in time for the annihilation gamma photons to be the correct frequency to carry off the energy equivalent of the input particles, the pre-annihilation proton / antiproton or electron / positron.

It would appear that in the case of the Big Bang mutual annihilation was much more difficult, and rare, than one might have assumed. And, thus it would appear that a large scale annihilation of matter and antimatter could not have taken place in the Big Bang with the result that the present universe must contain both matter and antimatter, most likely in equal amounts because of the original symmetry.

### 3 – *A Universe Containing Both Matter Regions and AntiMatter Regions*

Contemporary physics' position that there is now in the universe no antimatter from the Big Bang derives from the following reasoning.

If there are in the universe regions of antimatter as well as regions of matter, then at the boundaries between the two different type regions antimatter should come in contact with the matter. That should result in major amounts of mutual annihilations and the production of major amounts of gamma photons. Such major amounts of gamma photons not having been detected it is presumed that there are no antimatter regions of space.

Of course, an alternative explanation of such major amounts of gamma photons not having been detected would be that the mutual annihilations are not occurring in significant amounts for reasons similar to their not having so occurred in the original Big Bang or for other reasons. That alternative must now be investigated.

### *Why Matter and AntiMatter Regions Are Able to Co-Exist*

Of course, matter / antimatter mutual annihilations in general are not as awkward as they were for the original Big Bang with its peculiar initial conditions. Of interest here, however, is the case of the interstellar medium. It is the interstellar medium that must be examined because it is the natural boundary between regions of matter and regions of antimatter; where, if they are to occur, the anticipated matter / antimatter annihilations should be occurring and yielding their looked-for gamma ray flux.

In the interstellar [and intergalactic] medium the particles and antiparticles start from being significantly separated, residing in the vacuum of interstellar space, which vacuum, while not devoid of competing particles, has a much lower particle density than the original Big Bang. They do not suffer the disadvantage of being in a dense milieu of particles and antiparticles whose Coulomb attractions tend to cancel out their effects. And, they avoid the disadvantage of always starting their mutual Coulomb attraction toward each other with no initial velocity. Without regard for any mutual attraction between particular particles and antiparticles, they all move with significant velocities.

However, those velocities are in general not oriented toward the combination of a pair. Rather, the velocity directions are a combination of [a] some component distributed randomly over the particles in essentially all possible directions, and [b] some amount corresponding to a general flow direction.

Table 1, below summarizes the particle [and antiparticle where applicable] content of interstellar space. The density of the particles, and their related mean distance apart are such as to militate against any significant number of encounters, whether aided by Coulomb attraction or not. [Excepting solar wind, which is local to star's nearby environment, most of the interstellar medium is Hydrogen atoms, not ions.] [Gravitation can be ignored here, it being decades of orders of magnitude weaker than Coulomb attraction.]

| Region | Size | Particle | |
|---|---|---|---|
| | | Density [/cc] | Energy |
| Our Solar Wind | Sun Neighborhood | 10. | 0.001- 0.004 × c |
| Our Local Cloud | 60 Light Years | 0.1 | ~ 7,000 °K |
| Our Local Bubble | 300 Light Years | 0.001 | ~ 1,000,000 °K |
| Intergalactic Space | [The Universe] | 0.000 … ? | ? |

*Table 1 – The Interstellar Medium*

As has been pointed out in analyses of our solar wind, with typically `1 atom` in each `10 cm`$^3$ of interstellar gas in our local cloud and `10 ions` in each `cm`$^3$ of our solar wind, the particles are so far apart that the solar wind and interstellar gas flow through each other without being disturbed by collisions. On that basis, the even less dense regions of the interstellar medium such as ones like our local bubble, those within galaxies in general, and those in intergalactic space are even less conducive to particle / antiparticle encounters.

Another factor bearing on the likelihood of matter / antimatter mutual annihilations occurring in interstellar space is as follows. Because gravitational and Coulomb field attraction communicate at `c`, particles are attracted to where the attractor was, not where it is. That tends to produce orbital motion or "sling shot" non-collision passages rather than direct collisions. For example, a proton traveling at `0.000,001·c` `[only 300 meters/second]` and at a distance of `0.001 millimeter` from another charged particle [compare that distance with the spacing implied by the densities of the above table] will travel a distance equal to `757 of its proton radii` during the time that its Coulomb field communicates at velocity $c$ to the other charged particle its then Coulomb attraction impulse.

All of these various factors taken into account, matter / antimatter collisions must be quite infrequent events in the interstellar medium. When such mutual annihilations occur the appropriate gamma photons are emitted; however, the rate of occurrence of such events must be so low that nothing approximating a detectable extensive "gamma flux" could be produced.

### *Separation of Matter and AntiMatter In the Universe*

The original, spherically symmetrical Big Bang, in spite of its symmetry developed over a period of time into the present universe which has a substantial amount of non-symmetrical structure. The "clumping" that produced the current structure of various cosmic bodies would have operated equally on both matter and antimatter and on the two intermixed. The result must have been cosmic structures of various mixtures of matter and antimatter which developed into purely one or the other.

In regions of space where the particle density became greater, with young stars gradually assembling particles from their surroundings, there must have been a significant amount of matter / antimatter mutual annihilation. Depending on the local relative amounts of the two the resulting star would have developed into one of pure matter or of pure antimatter.

But the developed star, residing in its share of interstellar medium, would be able to continue its existence as a pure matter star or a pure antimatter star, essentially shielded or protected by its surrounding interstellar medium. In that manner the present universe should consist overall of equal amounts of matter and of antimatter, the two types residing in their pertinent stars, sufficiently "insulated" from each other by the vast and sufficiently empty interstellar medium such that their mutual annihilation with their matter-type opposite should be a relatively rare event.

### *Indications of Some Matter / AntiMatter Mutual Annihilations*

The most likely indication of our detection of cosmic matter / antimatter annihilations is Gamma Ray Bursts [GRB's] [3]. However, because of contemporary physics' conviction that the universe is now without antimatter that possibility has been left rejected and uninvestigated.

GRB's are the most luminous events known in the universe since the Big Bang. They are flashes of gamma rays coming from seemingly random places in deep space at random times. GRB's last from milliseconds to minutes, and are often followed by "afterglow" emission at longer wavelengths. Gamma-ray bursts are detected by orbiting [*Swift*] satellites about two to three times a week, as of 2007, though their actual rate of occurrence may be higher. All known GRB's come from outside our own galaxy. Most GRB's come from billions of light years away [as much as $z = 6.3$ or more].

Under the assumption that a given burst emits energy uniformly in all directions, some of the brightest bursts correspond to a total energy release of $10^{47}$ $joules$, nearly a solar mass converted into gamma-radiation in a small amount of time. No candidate process that has been considered to explain GRB's is able to liberate that much energy so quickly.

That energy requirement is eased if the burst is actually funneled out along a narrow jet with an angle of a few degrees. Then the actual energy release for a typical GRB is comparable to that of a very luminous supernova. For that reason the longer duration majority of observed GRB's are thought to be such collimated emissions.

They are hypothesized to come from the core-collapse of rapidly rotating, high-mass stars into black holes. Called *hypernovae*, those are stars that are sufficiently large to collapse directly into a black hole. If they indeed occur at all, which is not known, they would be quite rare. How the energy from that gamma-ray burst progenitor is turned into gamma radiation is a major topic of research. Neither the light curves nor the spectra of GRB's show resemblance to the radiation emitted by that candidate physical process.

Matter / antimatter annihilations are a promising source of GRB energy in a universe containing equal amounts of matter and antimatter because only such annihilations involve the total conversion of all of the participating mass into energy.

The actual observed energy of GRB's, from which the attribution of overall omni-directional energy comparable to a solar rest-mass is calculated, is itself compatible with multiple individual particle-antiparticle annihilations in all of which the gamma photons happen to be directed toward we the observers. The calculated extrapolation to the uniform omni-directional case would, then, be a calculation of the rate of actual separate individual particle / antiparticle annihilations with their gamma ray photons being randomly directed in all directions – the total mass conversion rate and that, on the average, for the overall entire universe to the extent that we are detecting all available GRB's. The so-determined mass conversion in on-going particle / antiparticle annihilations amounts to a minute portion of the universe's overall mass and is consistent with the rarity of such events as presented earlier above.

In a universe with regions of matter and antimatter "drifting" without annihilating in general but with a wide range of separation distances and conditions and a wide range of relative velocities some pairs of such regions are more likely to be readily accessible to each other for a matter / antimatter annihilation than others. The most accessible would be the earliest to annihilate. The "using up" of those leaves the remaining likelihood of annihilation for the others so reduced.

Those are the standard conditions for the usual exponential decay form. That is, the number, $n$, of occurrences of cosmic matter / antimatter annihilations would be expected to vary with time, $t$, as equation *(11)*,

*(11)* $$n = N_0 \varepsilon^{-t/\tau}$$

where $\tau$ is the (unknown at present) decay constant of the process and $N_0$ is the (also unknown at present) initial number.

In contemporary opinion, GRB's are generally thought to be the result of the death of massive stars. It has recently [2007] been reported [4] that the observed rate of GRB's increases with red shift [of the GRB source] $z$ over the range $z = 0 - 4$ approximately as $(1 + z)^{1.5}$. That is, the indication is that the rate increases significantly with time into the past, and with an increasing rate of increase, at least back to the time corresponding to $z = 4$ [and probably back to the Big Bang].

As presented in the report [4], that finding is inconsistent with the massive supernovae core collapse hypothesis for GRB's. On the other hand, it is entirely consistent with the expected variation with time of the number of occurrences of cosmic matter / antimatter annihilations as in equation *(11)*. Thus the findings in the report are consistent with the foregoing matter / antimatter annihilations analysis, the conclusion that the universe consists of equal amounts of matter and antimatter, and the hypothesis that the cause of GRB's is massive particle / antiparticle annihilations.

Per the data reported by the *Swift* satellite any particular observed GRB appears to come from a particular location in the universe however in general a different such location for each different GRB. The massive particle / antiparticle annihilations event at that location could be the collision of two stellar bodies or of two regions of relatively dense interstellar gas or of something else on that order.

In whichever case it is difficult to envision how the requisite total number of individual particle / antiparticle annihilations could take place in the requisite brief duration of the observed bursts. The best available analogical example is that of a nuclear fusion weapon's explosion in which at the extreme energies produced by its nuclear fission "trigger" and then by the on-going fusion explosion itself the requisite large number of intimate particle-particle interactions takes place in an extremely brief amount of time.

On the other hand, the violent encounter of two stellar bodies having a total mass comparable to a solar rest mass so taking place that they essentially merge and interact in a matter / antimatter annihilation in a time of a moderate number of minutes is not unreasonable. For example, at a joint approach velocity of $0.1 \cdot c$ they would travel together $30,000,000$ $meters$ in one second and $1,800,000,000$ $meters$ in one minute, a distance greater than a solar diameter [$1.4 \cdot 10^9$ $meters$] in a time typical of a GRB.

Such type events should be able to account for the entire range of GRB energies and durations from the very long to the short bursts. Such type events are to be expected in a universe of equal amounts of matter and antimatter "floating" in "islands" at various separations and velocities.

## *4 – Conclusion*

The current assumption, that an original skewed balance of the amounts of matter and antimatter produced in the Big Bang account for our universe having survived an original massive mutual annihilation of original matter and antimatter, is difficult to justify. Current investigations seeking to detect an innate violation of matter / antimatter symmetry sufficient to justify the original matter being greater in amount than the original antimatter have not had success. The concept is contrary to the natural condition that the Big Bang had to be smoothly spherically symmetrical in its particles of both matter and antimatter, its energy, and its radiation emitted outward from the origin.

The alternative proposed in the present paper, that no significant amount of matter / antimatter annihilations took place at the time of the spherically symmetrical Big Bang because it was physically impossible and that the universe consists of equal amounts of matter and antimatter, requires no questionable assumptions and is in comfortable agreement with known physics and astrophysics. It therefore merits consideration and acceptance in place of the skewed balance concept.

Similarly, matter / antimatter annihilations merit consideration and acceptance in place of the collimated black hole core collapse hypothesis for the origin of GRB's.